\documentclass[twocolumn]{aastex63}

\usepackage{graphicx}	
\usepackage{color}
\usepackage{amsmath}	
\usepackage{amssymb}	
\usepackage{mathrsfs}
\usepackage{mathrsfs}
\usepackage{multirow}

\newcommand{\msun}{M_\odot}

\newcommand{\mpc}{{\rm cMpc}}

\newcommand{\Muv}{M_{\rm UV}}

\newcommand{\beq}{\begin{equation}}
\newcommand{\eeq}{\end{equation}}

\shorttitle{A Soltan argument for LRDs}
\shortauthors{Inayoshi \& Ichikawa}

\begin{document}

\title{Birth of Rapidly Spinning, Overmassive Black Holes in the Early Universe}

\correspondingauthor{Kohei Inayoshi}
\email{inayoshi@pku.edu.cn}

\author[0000-0001-9840-4959]{Kohei Inayoshi}
\affiliation{Kavli Institute for Astronomy and Astrophysics, Peking University, Beijing 100871, China}

\author[0000-0002-4377-903X]{Kohei Ichikawa}
\affiliation{Global Center for Science and Engineering, Faculty of Science and Engineering, Waseda University, 3-4-1, Okubo, Shinjuku, Tokyo 169-8555, Japan}
\affiliation{
Department of Physics, School of Advanced Science and Engineering, Faculty of Science and Engineering, Waseda University, 3-4-1,
Okubo, Shinjuku, Tokyo 169-8555, Japan}

\begin{abstract}
The James Webb Space Telescope (JWST) has unveiled numerous massive black holes (BHs) in faint, broad-line active galactic nuclei (AGNs).
The discovery highlights the presence of dust-reddened AGN populations, referred to as ``little red dots (LRDs)",  
more abundant than X-ray selected AGNs, which are less  influenced by obscuration.
This finding indicates that the cosmic growth rate of BHs within this population does not decrease but rather increases at higher redshifts beyond $z\sim 6$. 
The BH accretion rate density deduced from their luminosity function is remarkably higher than that from other AGN surveys in X-ray and infrared bands.
To align the cumulative mass density accreted to BHs with the observed BH mass density at $z\simeq 4-5$,
as derived from the integration of the BH mass function, the radiative efficiency must be doubled from 
the canonical 10\% value, achieving significance beyond the $>3\sigma$ confidence level. 
This suggests the presence of rapid spins with 96\% of the maximum limit among these BHs under the thin-disk approximation, maintained by prolonged mass accretion instead of
chaotic accretion with randomly oriented inflows.
Moreover, we derive an upper bound for the stellar mass of galaxies hosting these LRDs, ensuring consistency with 
galaxy formation in the standard cosmological model, where the host stellar mass is limited by the available baryonic reservoir.
Our analysis gives a lower bound for the BH-to-galaxy mass ratio that exceeds the typical value known in the nearby universe
and aligns with that for JWST-detected unobscured AGNs.
Accordingly, we propose a hypothesis that the dense, dust-rich environments within LRDs facilitate the emergence of 
rapidly spinning and overmassive BH populations during the epoch of reionization.
This scenario predicts a potential association between relativistic jets and other high-energy phenomena with overmassive BHs in the early universe.

\end{abstract}
\keywords{Galaxy formation (595); High-redshift galaxies (734); Quasars (1319); Supermassive black holes (1663)}

\section{introduction}
The cosmic evolution of massive black hole (BH) populations is predominantly driven by mass accretion, powering active galactic nuclei 
(AGNs) \citep[e.g.,][]{Lynden-Bell_1969} with a certain level of merger contributions to BHs harbored in nearby massive ellipticals
\citep[e.g.,][]{McWilliams_2014,Kulier_2015}.
Multi-wavelength observations have consistently shown that AGN activity peaks around $z\sim 2$ and declines toward higher redshifts 
\citep[e.g.,][]{Ueda_2014, Delvecchio_2014}. 
The analysis of AGN activity offers insights into the radiative efficiency of accreting BHs by comparing it with the local mass density 
of relic BHs \citep{Soltan_1982,Yu_Tremaine_2002}.

\begin{figure*}
\begin{center}
{\includegraphics[width=86mm]{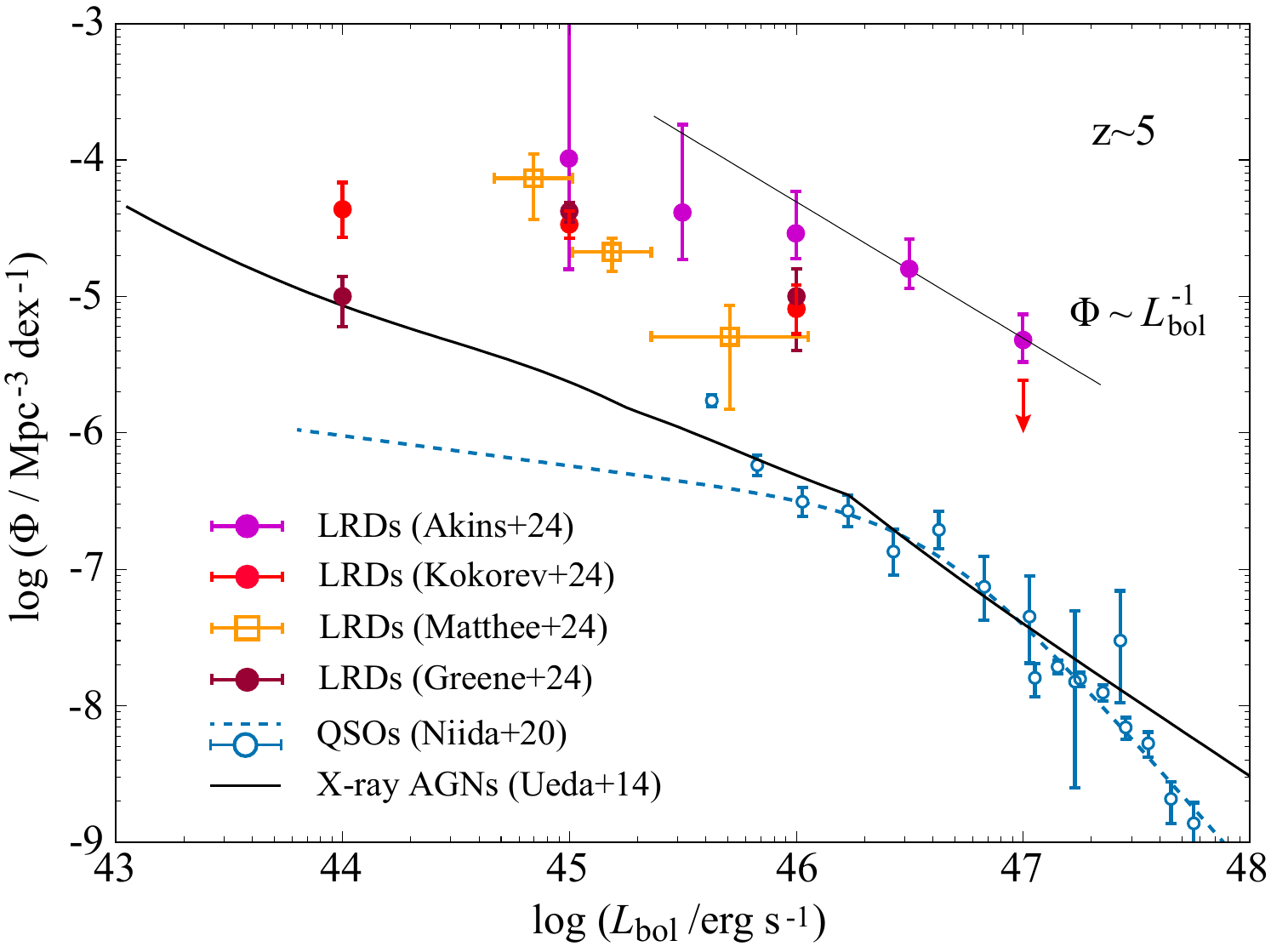}}\hspace{3mm}
{\includegraphics[width=86mm]{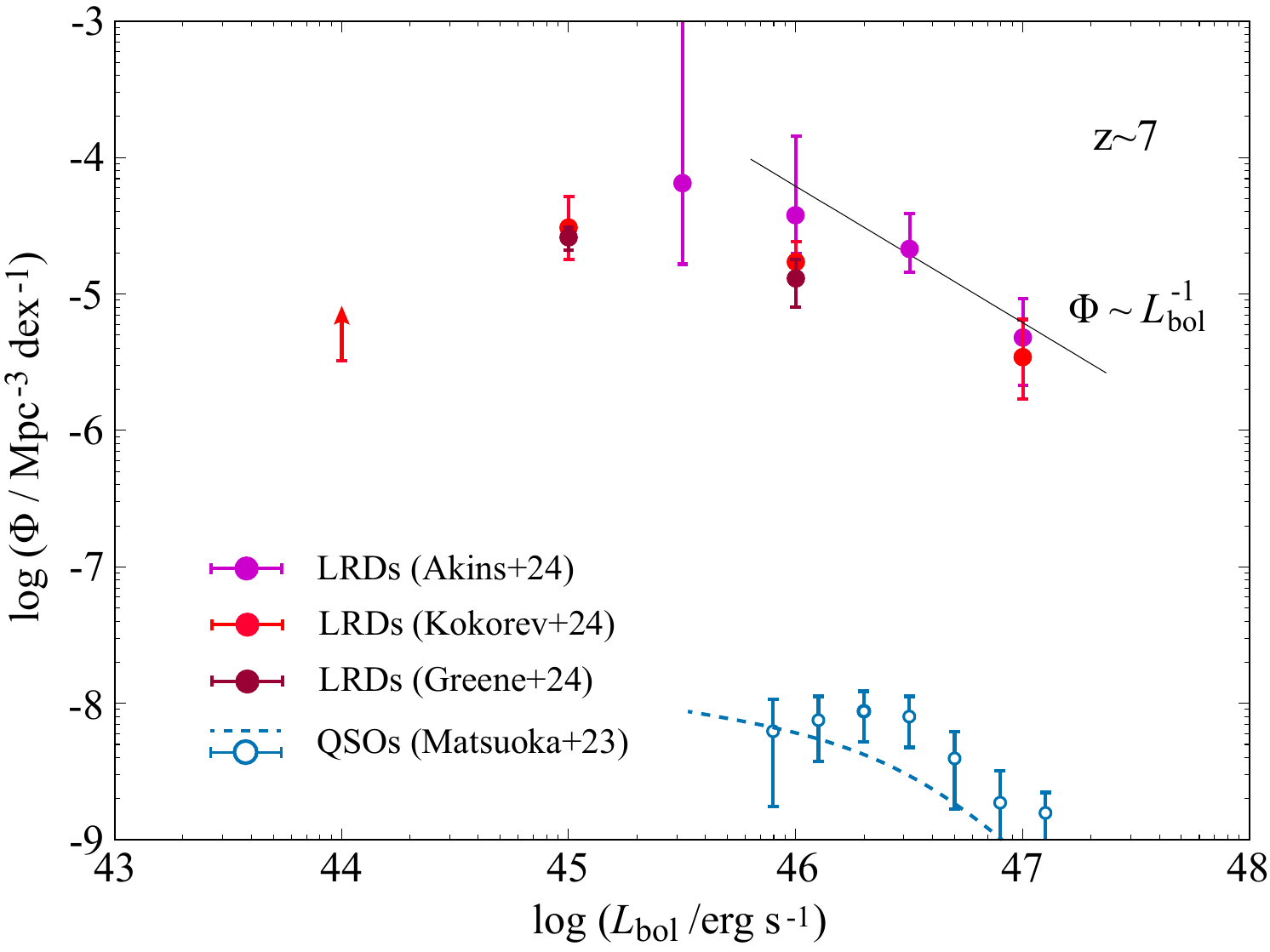}}
\caption{Bolometric AGN luminosity functions at $4.5<z<6$ (left) and $6.5<z<8.5$ (right). 
The luminosity function data obtained from different surveys are shown: the rest-UV-selected quasars \citep{Niida_2020,Matsuoka_2023},
the X-ray selected AGNs \citep{Ueda_2014}, and dust-reddened AGNs reported as ``little red dots (LRDs)" identified with JWST photometry and slitless spectroscopy 
\citep{Matthee_2024, Kokorev_2024, Greene_2024,Akins_2024}.
The bright end slope of the LRD luminosity function is consistent with $\Phi \propto L_{\rm bol}^{-1}$ at both redshifts.
}
\vspace{2mm}
\label{fig:QLF_bol}
\end{center}
\end{figure*}

Recent observations by the James Webb Space Telescope (JWST) have revealed a new category of dust-reddened, broad-line AGNs, often referred to as ``little red dots" \citep[hereafter LRDs,][]{Matthee_2024}. 
These AGNs are characterized by their compact morphology and moderate dust obscuration ($A_\mathrm{V}\approx3$) in the spectra 
\citep{Barro_2024,Kocevski_2023,Harikane_2023_agn,Labbe_2023}.
Investigations into the AGN luminosity function of LRDs at $z=4$--$8$ found an abundance of $\Phi \sim 10^{-5}-10^{-4}~\mpc^{-3}~{\rm mag}^{-1}$ in 
the observed UV absolute magnitude range of $-22\lesssim \Muv \lesssim -18$ \citep[e.g.,][]{Kokorev_2024},
significantly exceeding those predicted by extrapolations from the unobscured AGN luminosity function 
in ground-based surveys \citep[e.g.,][]{Niida_2020}.
Moreover, spectroscopic analysis of LRDs, facilitating direct measurement of broad H$\alpha$ emissions -- a tracer of AGN activity \citep{Greene_Ho_2005} --
has led to the construction of the AGN bolometric luminosity function for LRDs. 
This result suggests that in contrast to expectations based on AGN surveys in the pre-JWST era, the cosmic growth rate of BHs 
within this AGN population does not decline but appears to increase at higher redshifts ($z>6$).

In this {\it Letter}, compiling data from extensive high-$z$ AGN and LRD surveys,
we constrain the radiative efficiency of BHs in dominant LRD populations by comparing the observed BH mass density at $z\simeq 4-5$
to the mass accreted to BHs calculated from the BH growth rate, i.e., the Soltan-Paczy\'nski argument at the epoch of reionization.
This analysis suggests a radiative efficiency higher than the canonical 10\% value and favors rapid spins of these BHs 
under the thin-disk approximation \citep[e.g.,][]{Shakura_Sunyaev_1973,Novikov_Thorne_1973}.
Furthermore, we establish an upper limit for the stellar mass of galaxies harboring these LRDs, and give a lower bound for 
the BH-to-galaxy mass that exceeds the typical value known in the nearby universe and aligns with that for JWST-detected unobscured AGNs.
Consequently, we propose a hypothesis that the dust-rich environments in LRDs promote the emergence of rapidly spinning and overmassive BH populations.

Throughout this paper, we assume a flat $\Lambda$ cold dark matter (CDM) cosmology consistent with the constraints from Planck \citep{Planck_2020};
$h = 0.6732$, $\Omega_{\rm m}= 0.3158$, $\Omega_\Lambda = 1-\Omega_{\rm m}$, $\Omega_{\rm b}=0.04938$, and $\sigma_8 = 0.8102$.
It is important to note that observational studies referenced in our work have adopted a different set of cosmological parameters, $h = 0.7$ and $\Omega_{\rm m}= 0.3$.
However, the differences in parameter choice have a negligible impact on the results.

\section{Soltan Argument at $z\geq 5$}

Early studies of LRDs have indicated that the observed values of $\Muv$ are much fainter than the intrinsic UV magnitudes due to dust extinction
\citep[e.g.,][]{Kocevski_2023, Matthee_2024, Greene_2024, Kokorev_2024}. 
To address the underlying AGN activity, these studies estimated bolometric AGN luminosity from rest-optical emissions. 
\cite{Matthee_2024} and \cite{Greene_2024} conducted JWST/NIRSpec observations on LRDs and converted the measured H$\alpha$ luminosity to bolometric luminosity, 
while \cite{Kokorev_2024} and \cite{Akins_2024} relied on continuum luminosity $L_{5100}$ and $L_{3000}$ for a bolometric luminosity estimator, respectively, 
by assuming that all the continuum flux originates from the AGN.
The AGN bolometric luminosity derived from the H$\alpha$ luminosity, particularly its broad-line component emitted from fast-moving clouds 
near the AGN, is more accurate than using dust-dereddened continuum (see Section~\ref{sec:HatoBol}).

Figure~\ref{fig:QLF_bol} presents the bolometric luminosity functions at $z\simeq 5$ (left) and $z\sim 7$ (right),
i.e., the number density per unit comoving volume per $\log L_{\rm bol}$ in units of $\mpc^{-3}~{\rm dex}^{-1}$, 
combining data from various sources: 
rest-UV-selected unobscured quasars \citep{Niida_2020, Matsuoka_2023}, X-ray selected AGNs \citep{Ueda_2014}, 
and LRDs identified through JWST photometry and spectroscopy \citep{Matthee_2024, Kokorev_2024, Greene_2024,Akins_2024}\footnote{
\citet{Akins_2024} reported new LRD samples identified through the COSMOS-Web survey during the revision of our manuscript.
The wide survey areas allow the detection of rarer and brighter LRD populations at $5<z<9$, and thus better constrain the abundance at the bright end.}.
For UV and X-ray selected AGNs, we adopt the bolometric correction factors calibrated by \citet{Duras_2020}.
The abundance of unobscured quasars at $z\sim 5$ aligns closely with that of X-ray detected AGNs for $L_{\rm bol} \gtrsim 4 \times 10^{45}~{\rm erg~s}^{-1}$, 
while the X-ray AGN abundance further increases at lower luminosity regimes\footnote{\citet{Niida_2020} excluded the data 
at $\Muv > -23.32$ mag, corresponding to the two faintest data points in the left panel of Figure~\ref{fig:QLF_bol}, from their fitting
because the contamination rate of point-like compact galaxies increases significantly toward the fainter regime.}. 
This suggests that the obscured AGN fraction increases toward the fainter end \citep{Vito_2018}.
In contrast, LRDs exhibit a substantially greater abundance at $L_{\rm bol} \lesssim 10^{46}~{\rm erg~s}^{-1}$, nearly one order of magnitude 
above that of X-ray AGNs.
While the LRD abundance data show some variations owing to differences in sample size and methods for estimating bolometric luminosity 
across the referred studies, the trend of over-abundance relative to the other AGN populations is consistently observed.
The bright end slope of the LRD luminosity function is consistent with $\Phi \propto L_{\rm bol}^{-1}$ \citep{Kokorev_2024,Akins_2024}.
Therefore, it is ensured that a large fraction of the production of radiation (i.e., the amount of material accreted to BHs) is dominated by these bright populations.
The over-abundance of LRDs and the bright-end slope of $\Phi \propto L_{\rm bol}^{-1}$ hold even at $z\gtrsim 7$.

In our analysis below, we consider abundance data from luminosity bins where the sample size is $N\geq 2$, but exclude bins with a sample size of $N=1$.
This approach aligns with Poisson statistical error estimates, where a single occurrence ($N=1$) is statistically indistinguishable from zero.
Additionally, for the COSMOS-Web luminosity function data, we exclude the data at the faint end of $L_{\rm bol}<10^{46}~{\rm erg~s}^{-1}$ from \citet{Akins_2024}
since the COSMOS-Web survey is not deep enough to accurately measure the abundance of these faint populations and thus 
completeness correction matters.

\begin{figure}
\begin{center}
{\includegraphics[width=84mm]{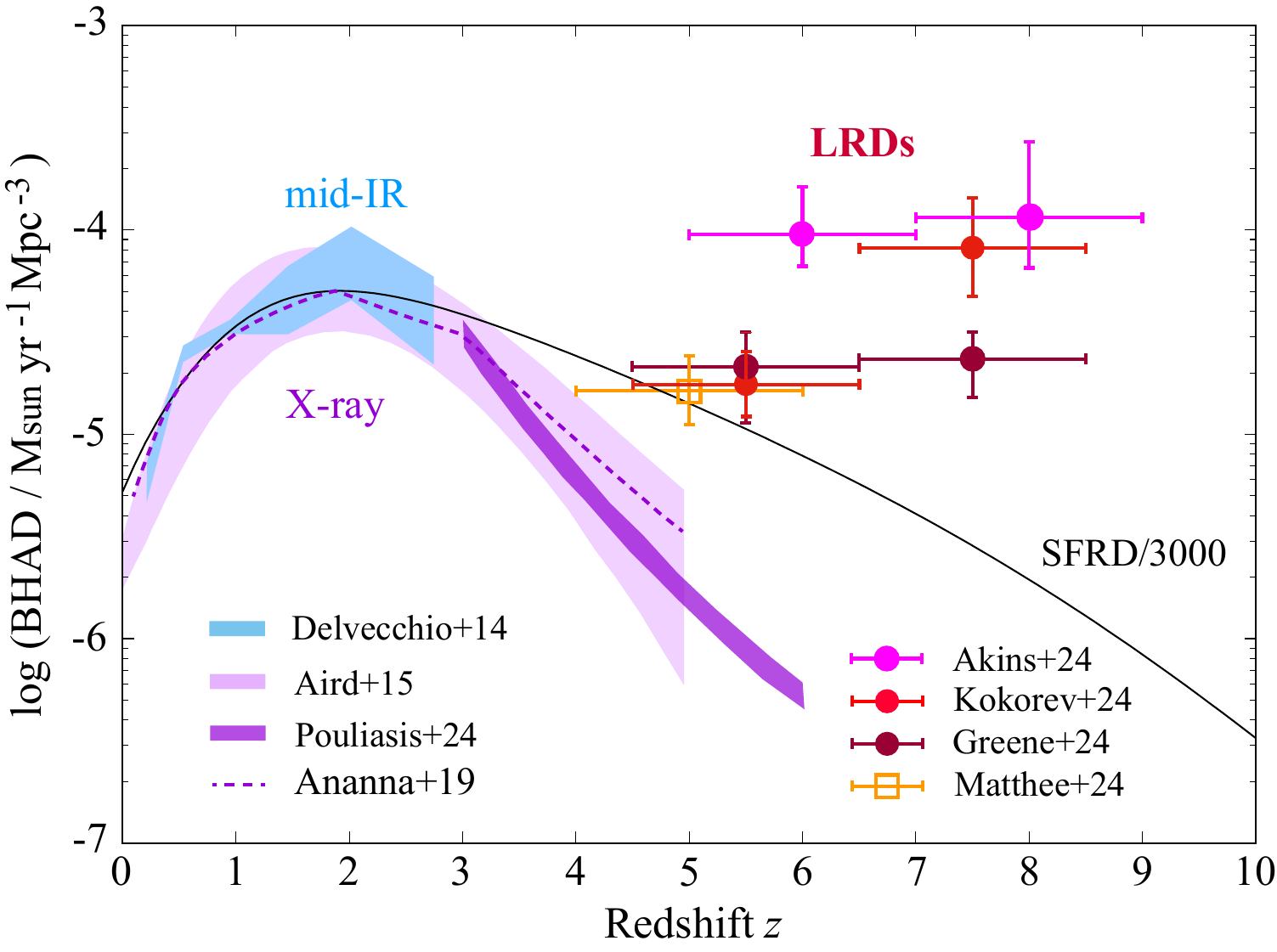}}
\caption{The cosmic BH accretion rate density (BHAD) as a function of redshift.
Each data point and curve represent BHADs estimated under the assumption of a 10\% radiative efficiency ($\epsilon_{\rm rad}=0.1$)
for the three different populations, including LRDs \citep{Matthee_2024, Kokorev_2024, Greene_2024,Akins_2024}, 
X-ray selected AGNs including Compton-thick populations \citep{Aird_2015,Ananna_2019,Pouliasis_2024}, and mid-infrared selected AGNs \citep{Delvecchio_2014}. 
For comparison, the cosmic SFRD scaled by a factor of 3,000 is overlaid \citep{Harikane_2022a}.
The BHAD attributed to LRDs remains significantly dominant at $z>6$.
}
\vspace{2mm}
\label{fig:BHAD}
\end{center}
\end{figure}

\begin{figure*}
\begin{center}
{\includegraphics[width=87mm]{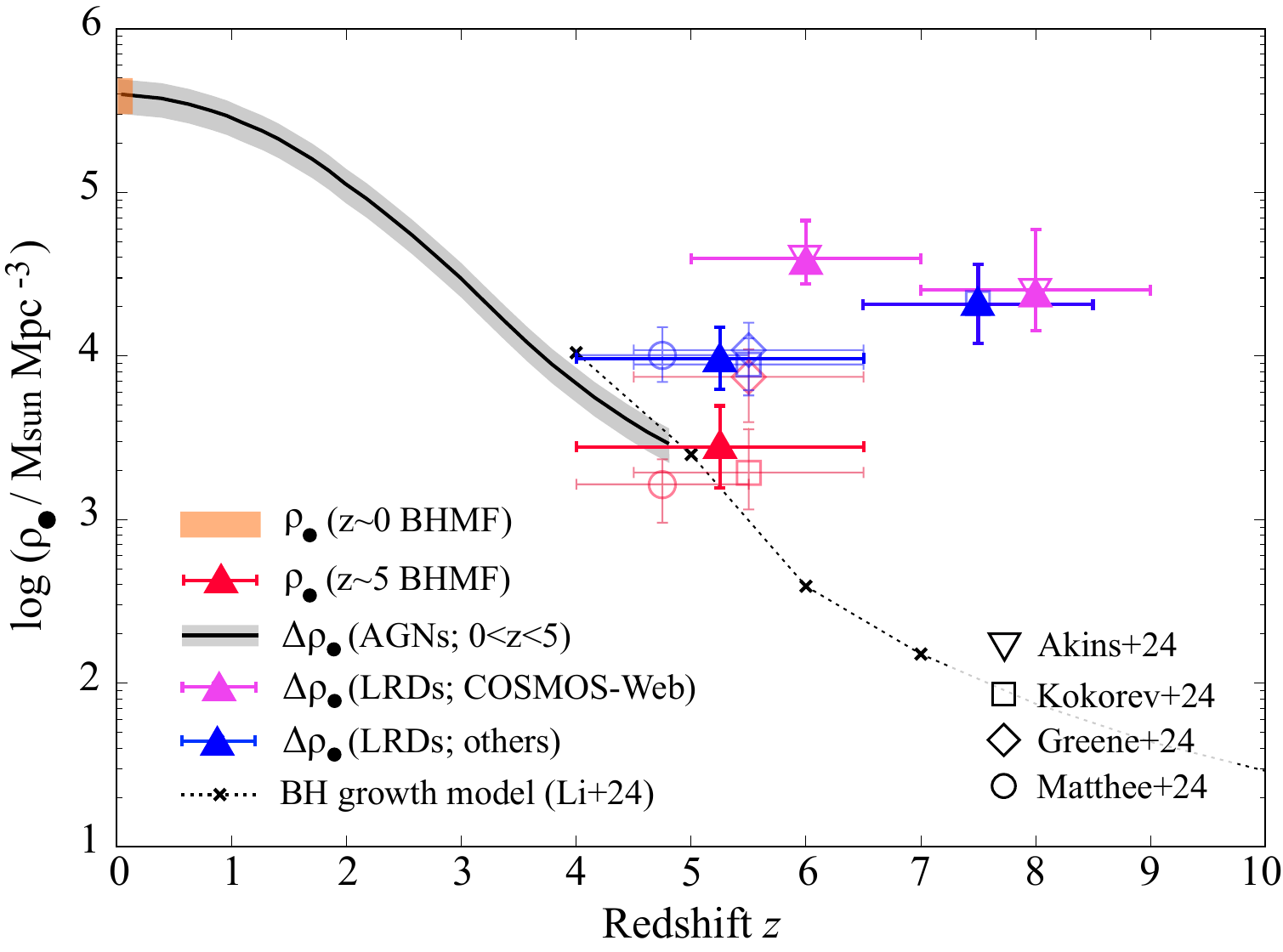}}\hspace{5mm}
{\includegraphics[width=80mm]{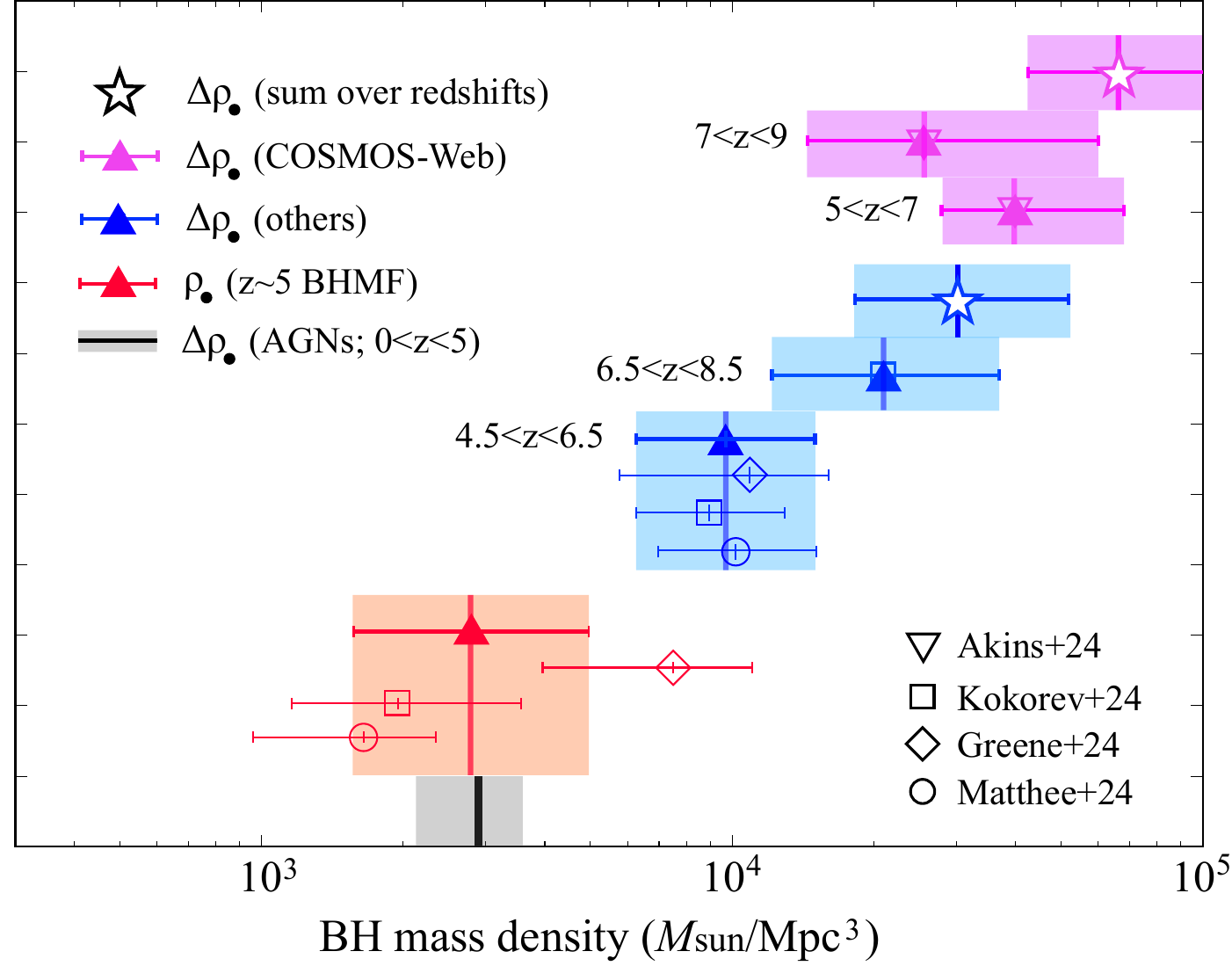}}
\caption{{\it Left}: Cosmic evolution of the BH mass density in a comoving volume. 
At $z\sim 5$, the BH mass density is derived from the integration of the BHMF for LRDs (red symbols). 
From this point, the mass density grows toward lower redshifts following the BHAD deduced from known 
AGN populations with a 10\% radiative efficiency (solid curve; \citealt{Ueda_2014}) and reaches the density of relic BHs 
in the nearby universe \citep{Shankar_2009}. 
At $z>5$, the cumulative mass accreted to BHs during the LRD phase, $\Delta \rho_\bullet \equiv {\rm BHAD}\times \Delta t$ 
inferred from their bolometric luminosity function over a time span $\Delta t$ for given redshift range based on the COSMOS-Web (magenta) and the other surveys (blue),
assuming a 10\% radiative efficiency, substantially exceed the observed mass density at $z\simeq 5$ as well as the predictions from 
a BH growth model calibrated with UV and X-ray selected AGN luminosity function \citep[dotted curve;][]{Li_2023b}. 
Data for LRDs are derived from luminosity functions and BH mass estimates provided in the literature
(open symbols, \citealt{Matthee_2024,Greene_2024,Kokorev_2024,Akins_2024}) and the mean values for each group (filled symbols).
{\it Right}: Summary of the BH mass density and the cumulative mass density during the LRD phase 
assuming a 10\% radiative efficiency. 
Shaded areas indicate the BH mass density $\rho_\bullet$ at $z\simeq 5$ (red) and the cumulative mass density accrued during 
the LRD stage calculated from the COSMOS-Web (magenta) and the other surveys (blue).
The total sum of $\Delta \rho_\bullet$ over the entire redshift range in each data group is shown with a star symbol.}
\vspace{2mm}
\label{fig:rhoBH}
\end{center}
\end{figure*}

Figure~\ref{fig:BHAD} illustrates the BH accretion rate density (BHAD) across various redshifts, with each data point and curve 
representing BHADs estimated under the assumption of a 10\% radiative efficiency ($\epsilon_{\rm rad}=0.1$). 
These include LRDs \citep{Matthee_2024, Kokorev_2024, Greene_2024,Akins_2024}, as well as X-ray selected AGNs 
\citep{Aird_2015,Ananna_2019,Pouliasis_2024}, and mid-infrared selected AGNs \citep{Delvecchio_2014}. 
The BHAD estimated from X-ray sources including Compton-thick AGN contributions agrees well to that of mid-infrared AGNs at $z\sim 3$.
With the same value of $\epsilon_{\rm rad}=0.1$, the BHAD inferred from the bolometric luminosity function of LRDs indicates
a persistent or even increasing trend toward higher redshifts ($4< z < 8.5$), opposite to the declining trend of X-ray selected AGNs.
Note that completeness corrections for some LRD samples we adopt \citep[e.g.,][]{Matthee_2024, Kokorev_2024} have not been fully implemented in estimating the abundance of faint sources, 
thereby potentially leading to an increased abundance at the faint sources.
Nevertheless, our finding is unlikely to alter because the brighter LRD populations dominantly contribute to the BHAD (i.e., ${\rm d}\log \Phi/{\rm d}\log L_{\rm bol}\simeq -1$).
Note that the BHAD estimated from the COSMOS-Web result increases by $<20\%$ when faint AGNs with $L_{\rm bol}<10^{46}~{\rm erg~s}^{-1}$ are included in our analysis.

For comparison, we overlay the cosmic star-formation rate density (SFRD) scaled by a factor of 3,000 \citep{Harikane_2022a}.
The scaled SFRD matches well the BHAD based on the mid-infrared selected AGNs at $z\lesssim 3$ and appears to be consistent with the BHAD of LRDs at $z\simeq 5-6$. 
On the other hand, the BHAD attributed to LRDs remains significantly dominant at $z>6$.
This finding, based on the assumption of $\epsilon_{\rm rad}=0.1$, indicates that rapid growth of BHs at these earlier epochs
established a trend of overmassive BH in terms of the $M_\bullet/M_\star$ ratio, 
as observed in recent JWST AGN studies at $z>6$ \citep[e.g.,][]{Harikane_2023_agn,Maiolino_2023_JADES,Pacucci_2023}.

\begin{table*}
\renewcommand\thetable{1} 
\caption{Significance of the difference between $\rho_\bullet(z\simeq 5)$ and $\Delta \rho_\bullet(z\gtrsim 5)$
for different radiative efficiencies.}
\begin{center}
\begin{tabular}{c|c|ccc|cccc}
\hline
\hline
Survey & Redshift & &$\log_{10} \Delta \rho_\bullet $ &&    \multicolumn{4}{c}{$p$-value} \\
 \hline
& && $\epsilon_{\rm rad}=0.1$ && $\epsilon_{\rm rad}=0.1$ & $\epsilon_{\rm rad}=0.2$ & $\epsilon_{\rm rad}=0.3$ & $\epsilon_{\rm rad}=0.42$\\
&& & $a_\bullet \simeq 0.674$ & & $a_\bullet \simeq 0.674$ &$a_\bullet \simeq 0.960$ & $a_\bullet \simeq 0.996$& $a_\bullet \simeq 1.00$\\
\hline 
COSMOS-Web& $5<z<9$ &&  $4.82 ^{+0.29}_{-0.19}$ && 0.00204 & 0.00569   &  0.0132 & 0.0350\\
Other surveys  & $4.5<z<8.5$ &&  $4.48^{+0.24}_{-0.22}$ &&  0.00291 & 0.0115   &  0.0378 & 0.151\\
\hline 
\end{tabular}
\label{tab:soltan}
\end{center}
\tablecomments{~Column (1): Survey. Column (2): Redshift ranges. Column (3): Cumulative mass density of BHs accreted during the LRD phase (in units of $\msun~\mpc^{-3}$) 
with a 10\% radiative efficiency.
Column (4)-(7): the $p$-value evaluated in the $t$-test for the null hypothesis between  
$\rho_{\bullet}(z=5)$ and $\Delta \rho_\bullet$ at $z>5$ for different values of the radiative efficiency (and the corresponding BH spin parameters).
Here, we consider two cases with LRD data based on the COSMOS-Web survey \citep{Akins_2024} and other LRD surveys 
\citep{Matthee_2024,Greene_2024,Kokorev_2024} for calculating the total cumulative mass.
The COSMOS-Web result requires $\epsilon_{\rm rad}\geq 0.2$ beyond the $>3\sigma$ confidence level,
while the confidence level is $\gtrsim 2\sigma$ with LRD samples from other surveys.
}
\end{table*}

In the left panel of Figure~\ref{fig:rhoBH}, we show the evolution of the BH mass density within a comoving volume throughout cosmic time.
The solid curve represents the cumulative BH mass density deduced from the AGN bolometric luminosity functions (primarily X-ray selected populations) at $0<z<5$, under an assumed 10\% radiative efficiency \citep{Ueda_2014}.
This projection agrees closely with the observed BH mass density at $z\simeq 0$ \citep{Shankar_2009}, concluding the plausibility of the pre-assumed radiative efficiency \citep[$\epsilon_{\rm rad}=0.1$;][]{Soltan_1982, Yu_Tremaine_2002}.
Additionally, we present the BH mass density directly derived from the integration of the BHMF for LRDs at $z\simeq 5$,
estimated to be $\rho_\bullet \simeq 2.8^{+2.2}_{-1.2}\times 10^3~\msun~\mpc^{-3}$
\citep[red symbols;][]{Matthee_2024,Greene_2024,Kokorev_2024}\footnote{We note that the BHMF from the LRD samples in \citet{Kokorev_2024} is 
derived from their AGN bolometric luminosity function, assuming an average Eddington ratio value 
$\langle \lambda_{\rm Edd}\rangle \simeq 0.3$, which is motivated by the samples of broad-line LRDs compiled by \citet{Greene_2024}.}.
This estimate remarkably aligns with expectations based on the Soltan argument assuming the conventional radiative 
efficiency of 10\% at $0<z<5$, further reinforcing the consistency across the three distinct physical measures.

Next, we extend our analysis to the universe at $z>5$, focusing on the cumulative mass of BHs accreted during the LRD phase, 
$\Delta \rho_\bullet \equiv {\rm BHAD}\times \Delta t$, calculated from their bolometric luminosity function over a redshift interval.
We categorize the LRD samples into two groups: those identified through the COSMOS-Web survey \citep[magenta,][]{Akins_2024} 
and LRDs from other observational programs \citep[blue,][]{Matthee_2024,Greene_2024,Kokorev_2024}. 
This classification is based on two considerations: (1) the redshift intervals different among the samples in the literature, requiring a uniform redshift bin size for comparative analysis; 
and (2) the need to evaluate the impact of wide-area surveys such as COSMOS-Web on the LRD Soltan argument.
As seen in the left panel of Figure~\ref{fig:rhoBH}, one can find the cumulative mass of BHs at $z > 5$ substantially 
exceeds the observed mass density at $z\simeq 5$ as well as the predictions from a BH growth model calibrated with UV and X-ray selected AGN luminosity functions 
\citep[dotted curve;][]{Li_2023b}. 
This discrepancy raises concerns about a potential violation of the BH mass conservation law; namely, $\rho_\bullet (z\simeq 5) \gtrsim \Delta \rho_\bullet(z>5)$ needs to be hold. 
Thus, the possible inapplicability of $\epsilon_{\rm rad}=0.1$ is suggested for the early universe beyond $z>5$.
Adjusting the radiative efficiency upwards impacts the inferred BHAD, which follows $\propto (1-\epsilon_{\rm rad})/\epsilon_{\rm rad}$.
For instance, adopting the theoretical upper limit of $\epsilon_{\rm rad}=0.42$ for an extreme Kerr BH with a spin parameter
$a_\bullet = 1$ \citep[e.g.,][]{Kerr_1963,Novikov_Thorne_1973} resolves the discrepancy between the integrated BHMF values and the cumulative mass derived from the BHAD. 
The relationship between the radiative efficiency and BH spin is well understood for geometrically-thin accretion disks, 
where a thermal equilibrium is maintained through efficient radiative cooling that balances with viscous heating \citep{Shakura_Sunyaev_1973}.
However, this scenario changes in low-accretion-rate states, where the disk becomes geometrically thick due to {\it inefficient} cooling \citep[e.g.,][]{Yuan_Narayan_2014}.
In such cases, the radiative efficiency substantially decreases from the values in the thin-disk approximation \citep{Inayoshi_2019}. 
This reduction in $\epsilon_{\rm rad}$ enlarges the discrepancy between $\rho_\bullet (z\simeq 5)$ and $\Delta \rho_\bullet(z>5)$,
rather than mitigating it.

The right panel of Figure~\ref{fig:rhoBH} provides a detailed quantitative comparison of BH mass density values from several studies,
using the least-squares method for fitting $\rho_\bullet$ and $\Delta \rho_\bullet$ at each redshift interval.
The cumulative values of $\Delta \rho_\bullet$ across the entire redshift range are denoted by star symbols for each LRD sample (see also Table~\ref{tab:soltan}).
For the case without the COSMOS-Web survey (blue symbols), the accreted mass density at $z\sim 5$ is found to be lower than that at $z\sim 7$.
This difference is primarily due to the finding of \cite{Kokorev_2024}, where $N=9$ luminous LRDs with $L_{\rm bol}\simeq 10^{47}~{\rm erg~s}^{-1}$
were identified at $6.5<z<8.5$ but only one was reported at $4.5<z<6.5$ within a large sample set of LRDs from multiple survey fields.
We note that luminosity function bins with a sample size of $N = 1$ are excluded in our analysis as a single occurrence is statistically indistinguishable from zero.
Therefore, the inclusion of these luminous populations substantially influences the BHAD estimate.
Using only the COSMOS-Web result (magenta symbols), we consistently observe higher values of $\Delta \rho_\bullet$ at the two redshift ranges, 
owing to the wide-area survey designed to identify more luminous and rarer populations.
As a result, the total sum in each case reaches as high as $\Delta \rho_\bullet \simeq 3.0^{+2.2}_{-1.2}\times 10^4$ and $6.6^{+6.3}_{-2.3}\times 10^4~\msun~\mpc^{-3}$, respectively (star symbol).
With the mean values, the cumulative mass densities during the LRD phases over $5<z<9$ appear to be $\gtrsim 10$ times higher than the BH mass density at $z\simeq 5$. 
However, there is a concern regarding the classification of both the LRD samples of \cite{Kokorev_2024} and \citet{Akins_2024},
where all photometrically selected LRDs are considered as AGNs due to the lack of spectroscopic observations (see also Section~\ref{sec:HatoBol}). 
Due to these concerns, the cumulative mass density of BHs and their difference from the BH mass density are considered to be upper bounds.

To understand the influence of each contribution of $\Delta \rho_\bullet$ on this analysis and the need for a radiative efficiency 
beyond the standard 10\% value, we explore two scenarios: one considering the contribution from the COSMOS-Web survey (magenta) 
and another one compiling LRD samples from other observational programs (blue).
In this work, to assess the statistical difference between $\rho_\bullet(z\simeq 5)$ and $\Delta \rho_\bullet$ for each of the two cases, 
we employ the $t$-test, which serves as an appropriate statistical method
to determine whether there is a statistically significant difference in the mean values between two groups with unequal sample variances.
The $p$-values, as summarized in Table~\ref{tab:soltan}, indicate that the hypothesis of agreement between the two quantities 
at $\epsilon_{\rm rad}=0.1$ is statistically rejected in both the two cases with a confidence level of $>99.7\%$ ($p<0.003$).
For the analysis with the COSMOS-Web survey result, a radiative efficiency greater than $\epsilon_{\rm rad}\geq 0.3$ is concluded with a confidence level of $>98\%$,
while the case with other LRD survey data suggest $\epsilon_{\rm rad}\geq 0.2$ with a similar confidence level.
This finding suggests that the majority of BHs within LRDs or similarly gas/dust-rich environments are likely to process rapid spins
with an average $\epsilon_{\rm rad}\geq 0.2-0.3$ (the corresponding BH spin is $a_\bullet \simeq 0.96-0.996$), indicating a prevalent condition 
of rapid angular momentum in BH growing environments in the early universe.

This result suggests that BH growth at these high redshifts, especially in LRDs, is likely dominated by 
prolonged accretion episodes with coherent angular momentum directions or a modest degree of anisotropy \citep[e.g.,][]{Volonteri_2005,Dotti_2013}, 
unlike short-lived chaotic accretion with randomly oriented inflows that tend to spin BHs down \citep[e.g.,][]{King_2008}.
Our conclusion on rapid spins of the early BH population will be directly testable through future 
gravitational-wave observations with the space-based detectors such as LISA, TianQin, and Taiji 
\citep[e.g.,][]{Amaro-Seoane_2023,Torres-Orjuela_2024}.

Intriguingly, clustering analyses of quasars and galaxies at $z\gtrsim 6$ suggest that the duty cycle of UV-bright quasars is as low as $\lesssim 1\%$ \citep{Eilers_2024,Pizzati_2024}, 
corresponding to a quasar lifetime of $1-10$ Myr, which is significantly shorter than the $e$-folding time assuming the Eddington accretion rate. 
This finding implies that most of the BH mass growth would have occurred in highly (UV-)obscured environments
and/or through episodic super-Eddington phases with a lower radiative efficiency (\citealt{Davies_2019}; see also \citealt{Inayoshi_2016,Inayoshi_2022}).
The first implication is consistent with the hypothesis that LRDs are moderately obscured AGNs \citep[e.g.,][]{Li_2024}.
The second implication, concerning the potential mass contribution from radiatively inefficient super-Eddington growth, would be constrained by our findings in this work and is 
left for future investigations.

\begin{figure}
\begin{center}
{\includegraphics[width=82mm]{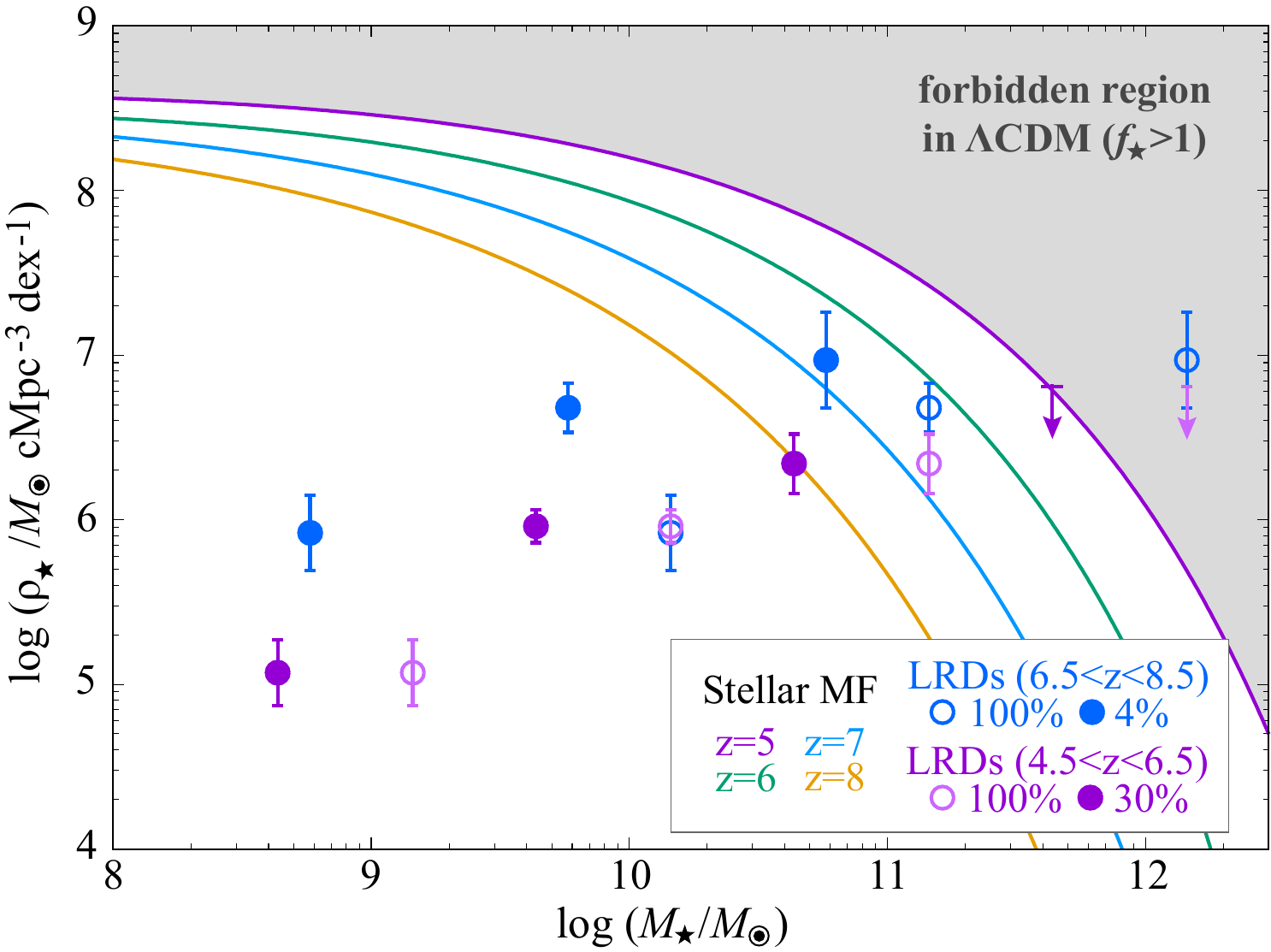}}
\caption{
Stellar mass density in galaxies hosting LRDs at various redshifts, calculated using Eq.~(\ref{eq:Mstar}) and assuming 
$\mathscr{F}(\equiv f_{\rm IMF}f_L)=1.0$ (open symbols) at two redshift ranges of $4.5<z<6.5$ and $6.5<z<8.5$.
For comparison, the stellar mass function derived from the DM halo mass function at $5\leq z \leq 8$ is shown
with a 100\% star formation efficiency.
An upper bound of the stellar mass constrains $\mathscr{F}<0.3 f_\star$ for $4.5<z<6.5$ and $\mathscr{F}<0.04 f_\star$ for $6.5<z<8.5$
(filled symbols).
}
\vspace{2mm}
\label{fig:starMF}
\end{center}
\end{figure}

\section{Potential overmassive BH trends in Little Red Dots}

In this section, we explore the possibility that BHs within LRDs are overmassive relative to the mass correlation with 
their host mass, as implied from the BHAD-to-SFRD ratio shown in Figure~\ref{fig:BHAD}.
In general, estimating the stellar mass of dust-obscured sources poses a significant challenge in the absence of 
rest-frame near-infrared data provided by JWST MIRI \citep[e.g.,][]{Williams_2024,Perez-Gonzalez_2024}.
Instead of examining the detailed spectral energy distribution fitting analysis, we focus on putting an upper bound for the stellar mass.
This approach ensures that the observed abundance of LRDs does not exceed the theoretical upper bound in the standard 
$\Lambda$CDM model with a 100\% conversion efficiency from gas to stars \citep[e.g.,][]{Boylan-Kolchin_2023}.

The stellar continuum for LRDs can be constrained by the dust-corrected continuum flux at 5100~{\AA}. 
Given that broad H$\alpha$ emission indicates AGN dominance in the continuum (see Section~\ref{sec:HatoBol}), 
we adjust $L_{\star,5100}=f_L L_{5100}$, where $f_L$ is significantly less than unity.
To estimate an upper bound of stellar mass, we employ the {\tt STARBURST99} population synthesis code 
\citep[version 7.0.1;][]{Leitherer_1999}, adopting a Kroupa IMF (\citealt{Kroupa_2001}; $0.1-100~\msun$), 
Padova isochrone models, constant star formation, and solar metallicity. 
This approach yields a galaxy mass$-$luminosity relation:
\begin{equation}
\frac{M_\star}{10^9~\msun} \simeq 1.3~f_{\rm IMF}\left(\frac{L_{\star,5100}}{10^{43}~{\rm erg~s}^{-1}}\right) \left(\frac{t_{\rm age}}{1~{\rm Gyr}}\right), 
\label{eq:Mstar}
\end{equation}
which is applicable for stellar ages of $t_{\rm age}\sim 0.3-3~{\rm Gyr}$.
This estimate is sensitive to the low-mass end of the stellar IMF ($f_{\rm IMF}=1$ for $m_{\rm \star,min}=0.1~\msun$); 
for instance, setting the minimum mass up to $m_{\star, \rm min}=1.0~\msun$ decreases the factor to $f_{\rm IMF}\sim 0.3$.

\begin{figure}
\begin{center}
{\includegraphics[width=80mm]{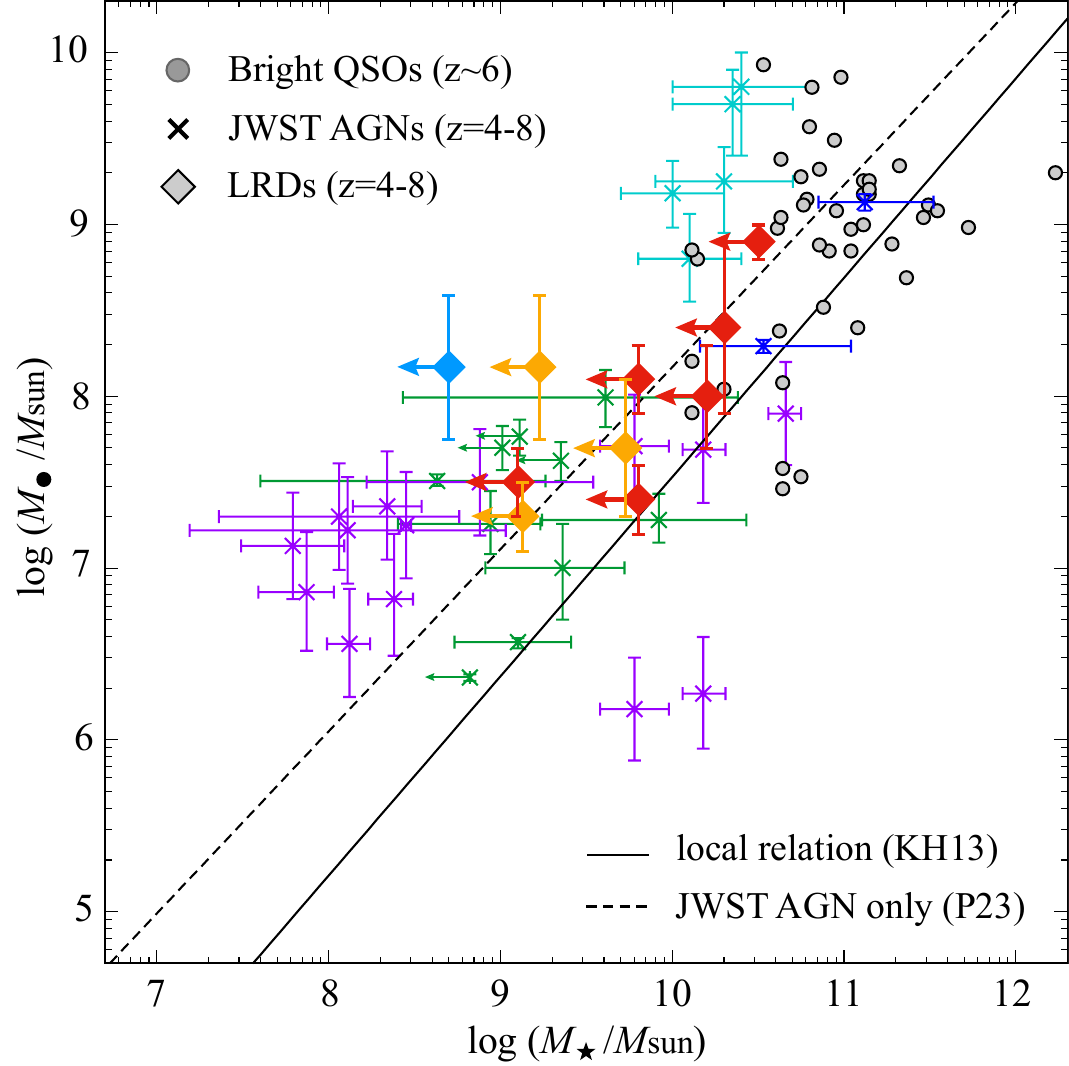}}
\caption{$M_\bullet - M_\star$ distribution for high-redshift AGNs, including LRDs,
JWST-detected unobscured ANGs at $z=4-8$ (purple, \citealt{Maiolino_2023_JADES}; green, \citealt{Harikane_2023_agn}; cyan, \citealt{Stone_2024};
and blue, \citealt{Ding_2023}),
and quasars identified in ground-based surveys \citep{Izumi_2021}.
For the LRDs at $4.5<z<6.5$ (red) and $6.5<z<8.5$ (orange), we derive the upper bound of the stellar mass based on the dust-corrected continuum 
flux measured by \cite{Greene_2024} using Eqs.~(\ref{eq:Mstar}) and (\ref{eq:FF}).
Additionally, a $z=8.5$ LRD with broad H$\beta$ emission, for which the stellar mass is constrained by ALMA non-detections, 
is overlaid \citep[blue,][]{Kokorev_2023}.
Two different mass correlations are overlaid: 
the local relationship \citep[solid,][]{Kormendy_Ho_2013} and the JWST-detected AGNs \citep[dashed,][]{Pacucci_2023}.
}
\vspace{-3mm}
\label{fig:Msigma}
\end{center}
\end{figure}

Figure~\ref{fig:starMF} presents the stellar mass function (in units of $\msun~\mpc^{-3}$) in galaxies hosting LRDs at various redshifts, 
calculated by using Eq.~(\ref{eq:Mstar}) and assuming $\mathscr{F}(\equiv f_{\rm IMF}f_L)=1.0$
at two redshift ranges of $4.5<z<6.5$ and $6.5<z<8.5$ (open circles), based on the LRD luminosity function obtained by \cite{Kokorev_2024}. 
Since the luminosity function of LRDs follows $\Phi \propto L_{\rm bol}^{-1}$, the stellar mass density becomes flatter at the high-mass end
when the stellar mass is translated from the luminosity with Eq.~(\ref{eq:Mstar}).
We compare the results to the stellar mass function derived from the halo mass function at $5\leq z \leq 8$, 
assuming $M_\star =f_\star f_{\rm b}M_{\rm h}$, where $f_{\rm b}=0.16$ is the cosmic baryon fraction and 
$f_\star$ is the star formation efficiency \citep[see more details in ][]{Inayoshi_2022c}. 
Setting $f_\star=1.0$ offers a theoretical upper limit on stellar mass in galaxies ($f_\star>1$; the forbidden region), 
highlighting a mismatch between the stellar mass density contained in LRDs with $\mathscr{F}=1.0$ and the $\Lambda$CDM upper limit.
As a result, we deduce a stringent constraint denoted with filled circles,
\begin{align}
\mathscr{F}<0.3~f_\star  & \hspace{2mm} {\rm at}~~4.5<z<6.5,\nonumber\\ 
\mathscr{F}<0.04~f_\star & \hspace{2mm} {\rm at}~~6.5<z<8.5.
\label{eq:FF}
\end{align}

Figure~\ref{fig:Msigma} shows the $M_\bullet - M_\star$ distribution for high-redshift AGNs, including LRDs (square),
JWST-detected unobscured AGNs (cross), and quasars identified in ground-based surveys (circle).
For the LRDs at $4.5<z<6.5$ (red) and $6.5<z<8.5$ (orange), we derive the stellar mass of those with broad H$\alpha$ emission 
from \cite{Greene_2024}, using Eq.~(\ref{eq:Mstar}).
This calculation incorporates the upper limit for the stellar continuum ratio $\mathscr{F}$ to the observed continuum 
(see Eq.~\ref{eq:FF} and Figure~\ref{fig:starMF}), 
ensuring consistency with the theoretical upper bound of stellar mass density in the $\Lambda$CDM universe.
The $M_\bullet - M_\star$ values for LRDs tend to be overmassive compared to the local relationship \cite[solid line;][]{Kormendy_Ho_2013},
and aligns well with other AGNs detected by JWST and follows the mass correlation inferred from JWST AGN data,
excluding quasars from ground-based surveys \citep[dashed line;][]{Pacucci_2023}.
Moreover, their distribution is consistent with the locus of a JWST/NIRSpec confirmed $z=8.5$ LRD that exhibits broad H$\beta$ emission, 
for which an upper limit on the stellar mass has been constrained by non-detection in ALMA observations \citep[blue;][]{Kokorev_2023}.

The mass ratios for these LRDs are also consistent with the BHAD/SFRD values at $z>6$ shown in Figure~\ref{fig:BHAD}.
This suggests a model in which transient rapid growth phases during the LRD stages elevate these BHs into an overmassive state.
This hypothesis is supported both theoretically \citep[e.g.,][]{Inayoshi_2022,Hu_2022b} and observationally \citep{Fujimoto_2022},
providing a comprehensive insight into the BH growth dynamics in the early universe.

\section{Discussion}
\subsection{Missing X-ray radiation from LRDs}

X-ray AGN surveys are generally effective in identifying obscured AGNs.
However, in the case of LRDs observed with JWST, no X-ray counterparts have been reported in the early studies \citep[e.g.,][]{Kocevski_2023,Furtak_2023a,Matthee_2024}.
X-ray weakness has been consistently observed in LRDs as demonstrated by stacking analyses \citep{Yue_2024},
and this phenomenon extends beyond LRDs to a more general category of unobscured broad-line AGNs \citep{Maiolino_2024}.
Further emphasizing the rarity of X-ray emissions, \citet{Kocevski_2024} have identified only two X-ray detected LRDs at $z=3.1$ and $4.66$ among 341 examined objects, 
resulting in a detection fraction of less than $0.6~\%$.
The optical continuum extinction measurements suggest a gas column density of $N_{\rm H} \sim 3.3 \times 10^{22}(A_{\rm V}/3.0)~{\rm cm}^{-2}$ 
\citep{Maiolino_2001}, indicating that the column density outside the broad-line region is too low to obscure X-rays.
The column density estimate is broadly consistent with those measured from the X-ray spectral analysis for the two X-ray detected LRDs,
$N_{\rm H} \sim (5-20) \times 10^{22}~{\rm cm}^{-2}$ \citep{Kocevski_2024}.

\begin{figure}
\begin{center}
{\includegraphics[width=84mm]{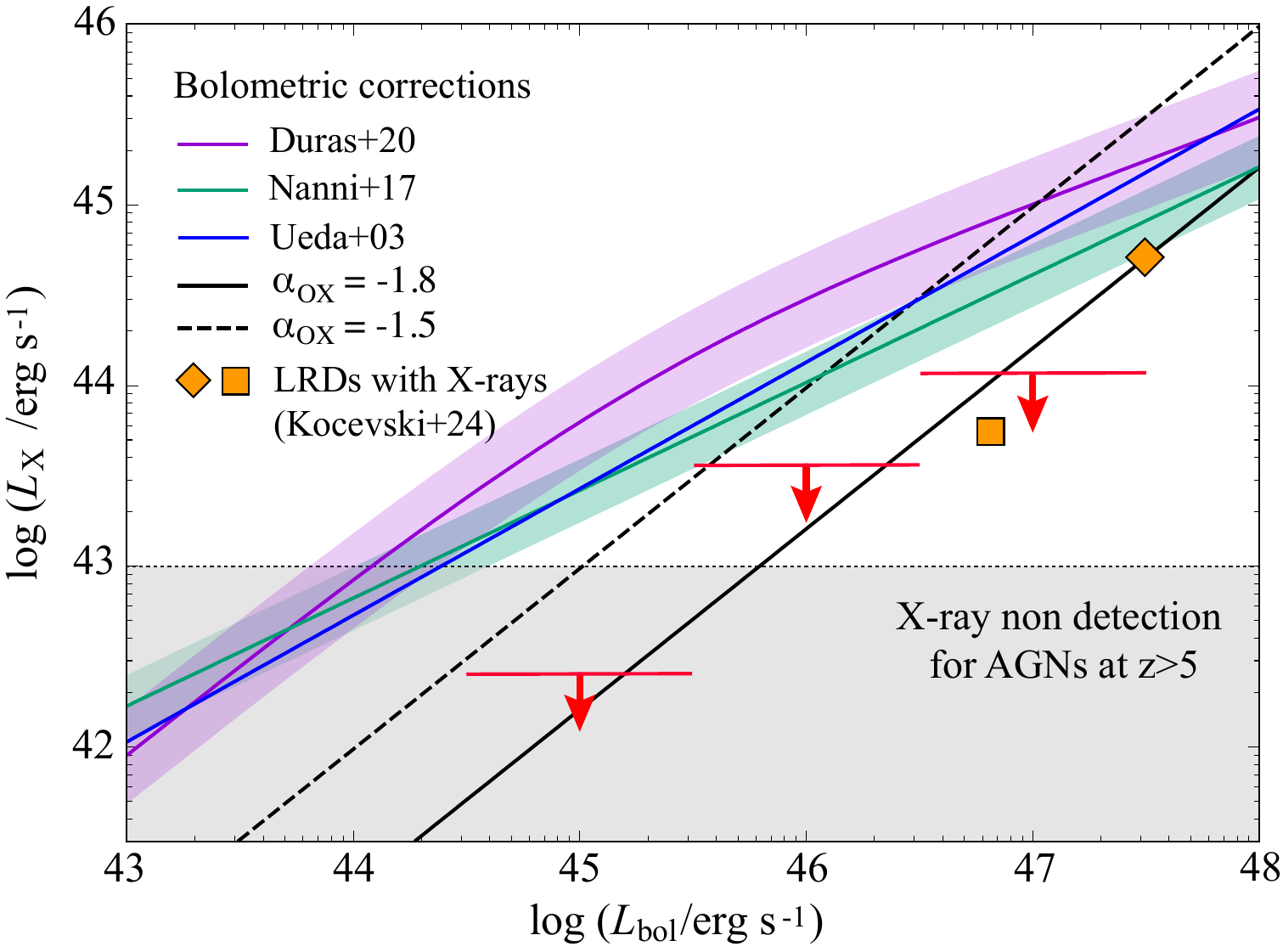}}
\caption{The critical X-ray luminosities for LRDs at $z\sim 5$, determined by the requirement for LRDs not to be classified as X-ray AGNs nor 
contributing to the abundance of X-ray AGNs (red horizontal lines with arrows).
For comparison, several models for bolometric correction to X-rays are shown \citep{Ueda_2003,Nanni_2017, Duras_2020}, 
as well as the cases with constant optical to X-ray spectral indices of $\alpha_{\rm OX}=-1.5$ (dashed) and $-1.8$ (solid).
The $L_{\rm bol}-L_{\rm X}$ values of the two LRDs detected in X-rays are shown with orange symbols; 
JADES 21925 (square) and PRIMER-COS 3982 (diamond) \citep{Kocevski_2024}.
The shaded area denotes X-ray luminosities below the detection threshold of current Chandra observations for JWST fields.
}
\vspace{-3mm}
\label{fig:ox}
\end{center}
\end{figure}

Considering the absence of X-ray counterparts for LRDs, we explore the possibility that their X-ray emission 
is intrinsically weak, as compared to typical X-ray selected AGNs. 
Figure~\ref{fig:ox} presents the critical X-ray luminosity for each bolometric luminosity, so that 
$\Phi_{\rm X}(L_{\rm X,crit}) \geq \Phi_{\rm LRD}(L_{\rm bol})$,
where we adopt the X-ray AGN luminosity function $\Phi_{\rm X}$ from \cite{Ueda_2014}
and the LRD bolometric luminosity function $\Phi_{\rm LRD}$ from \cite{Kokorev_2024}, respectively.
This condition requires that LRDs must have a bolometric correction factor to X-rays that prevents them from being classified as X-ray AGNs and 
contributing to their abundance (red horizontal lines with arrows).
The critical X-ray luminosity is limited below those derived from comparison between the X-ray and optical-based AGN luminosity functions at lower redshifts of
$z\lesssim 2$ \citep{Ueda_2003}.

To quantify this intrinsic X-ray faintness in LRDs, we utilize the optical to X-ray spectral index, 
defined as $\alpha_{\rm OX}=\log(L_{\rm \nu, 2keV}/L_{\nu, 2500})/\log(\nu_{\rm 2keV}/\nu_{2500})$,
where $L_{\rm \nu, 2keV}$ and $L_{\nu, 2500}$ are the extinction corrected luminosity density at $2~{\rm keV}$ and $2500~{\rm \AA}$.
Our findings support a constant value of $\alpha_{\rm OX}$ lower than $-1.8$, rather than the luminosity-dependent $\alpha_{\rm OX}$ values
observed in unobscured quasars across $0\lesssim z \lesssim 6$ \citep[e.g.,][]{Steffen_2006, Duras_2020}. 
The upper bound of $\alpha_{\rm OX}\simeq -1.8$ is seen in the most luminous quasar populations 
with $L_{\rm bol}\gtrsim 10^{47}~{\rm erg~s}^{-1}$ \citep{Nanni_2017}, which does not apply to most LRDs studied in this paper.

The extensive LRD samples by \cite{Kokorev_2024}, consisting of 260 dust-reddened AGN candidates, are compiled from
deep JWST/NIRCam fields totaling $\sim 340~{\rm arcmin}^2$.
This includes observations from the CEERS field, which spans $\sim 51.9~{\rm arcmin}^2$ and falls within the coverage area of 
the Chandra AEGIS-XD survey coverage \citep{Nandra_2015}. 
The survey detection limit reaches $1.5\times 10^{-16}~{\rm erg~s}^{-2}~{\rm cm}^{-1}$ at 0.5--10 keV,
which corresponds to $L_{\rm X}\simeq 1.1\times 10^{43}~{\rm erg~s}^{-1}$ at the rest-frame 2--10 keV for $z\sim 5-7$ sources
assuming a photon index of $\Gamma=-1.7$ and a Compton-thin limit ($N_\mathrm{H}<10^{24}$~cm$^{-2}$).
Given the average surface density of these LRDs ($\simeq 0.77~{\rm arcmin}^{-2}$), $\sim 40$ LRDs in the CEERS field show no X-ray detection above this threshold. 
This suggests that non-detection of X-rays among LRDs can be explained by the intrinsic faintness of X-rays as shown in Figure~\ref{fig:ox}.
Nevertheless, the most luminous LRDs with $L_{\rm bol}\gtrsim 10^{47}~{\rm erg~s}^{-1}$ might still be observed in X-rays 
unless classified as Compton-thick AGNs.
Additionally, we note that the two LRDs detected in X-rays with a modest hydrogen column density of $N_{\rm H}\sim 10^{23}~{\rm cm}^{-2}$, 
show obscuration-corrected X-ray luminosities of $L_{\rm X}\simeq 5.4 \times 10^{43}~{\rm erg~s}^{-1}$ (for JADES 21925 at $z_{\rm photo}=3.1$) 
$L_{\rm X}\simeq 5.0 \times 10^{44}~{\rm erg~s}^{-1}$ (for PRIMER-COS 3982 at $z_{\rm spec}=4.66$), respectively \citep{Kocevski_2024}.
The bolometric luminosities calculated from the rest-optical fluxes are $L_{\rm bol}\simeq 6.5\times 10^{46}~{\rm erg~s}^{-1}$ and 
$3.1\times 10^{47}~{\rm erg~s}^{-1}$ after dust attenuation correction.
These findings also suggest a lower value of $\alpha_{\rm OX}\simeq -1.8$ as shown in Figure~\ref{fig:ox} (orange symbols).

Alternatively, if the X-ray emission was not intrinsically faint (except the two X-ray detected LRDs), most of these LRDs would be embedded in Compton-thick gas 
with $N_\mathrm{H} \gg 10^{24}$ cm$^{-2}$, concealing the LRDs from deep X-ray observations.
However, such high densities are not expected in the LRD rest-optical spectra, which show a modest extinction $A_\mathrm{V} \sim 3$ mag, 
equivalent to $N_\mathrm{H} \sim 3 \times 10^{22}$ cm$^{-2}$.
This discrepancy is also consistent with the lack of expected absorption features in their NIRSpec rest-frame UV spectra of LRDs with broad-emission lines \citep[e.g.,][]{Greene_2024}.

\subsection{Dusty young starburst galaxies mimicking LRD-like AGNs?}

The classification of LRDs as AGNs has relied on the detection of broad H$\alpha$ emission.
However, such high-velocity gas can also originate from stellar processes, such as stellar winds or supernova explosions.
Wolf-Rayet (WR) galaxies, characterized by young and massive stellar populations, can exhibit broad H$\alpha$ emission as well as 
other high-ionization lines \citep[e.g.,][]{Ho_1995,Schaerer_1999}.
Therefore, to conclusively confirm the AGN nature of LRDs, it is essential to perform line diagnostics that extend beyond simply identifying broad H$\alpha$ emission,
e.g., detection of broad \ion{He}{2} $\lambda 4686$ emission, a signature frequently associated with WR galaxies
\citep[e.g., NGC 4214 discussed in][]{Sargent_Filippenko_1991}.

Spectroscopic observations provide valuable insights into the characteristics of emission lines in LRDs. 
Initial studies by \cite{Kocevski_2023} and \cite{Greene_2024} have found that the \ion{He}{2} $\lambda 4686$ line, 
commonly associated with WR stellar activity, is absent in these LRD sources. 
Additionally, the spectral signatures typically linked to WR stars have not been observed within the LRD samples.
The lack of \ion{He}{2} and other WR indicative spectral lines suggests that the broad H$\alpha$ emissions detected in LRDs are not of stellar origin
(note that detection of \ion{He}{2} $\lambda 4686$ does not necessarily exclude the AGN possibility because the emission is also observed in AGNs). 
This finding supports the hypothesis that AGNs are responsible for these emissions.
More detailed spectroscopic analyses of multiple emission lines would strength this conclusion \citep{Greene_ARAA_2020,Reines_2022}.

\subsection{Bolometric luminosity estimates}\label{sec:HatoBol}

Our study is motivated by intriguing discoveries on the high abundance of LRDs, but an important consideration about the AGN luminosity estimate needs to be noted.
While the UV luminosity functions show similar shapes across different studies, significant variations in the bolometric luminosity 
function arise due to the different methods used for bolometric luminosity estimates (see Figure~\ref{fig:QLF_bol}). 
Since the observed rest-frame UV flux is heavily attenuated, earlier studies have employed either the rest-frame continuum flux at 5100~\AA\ \citep{Kokorev_2024}, 
or the direct H$\alpha$ emission luminosity, when available, as a proxy for the AGN bolometric luminosity \citep{Matthee_2024,Greene_2024}.

The approach that relies on continuum flux introduces uncertainties of determining the AGN contribution to the total flux, which might result in overestimated luminosity. 
Furthermore, the selection of LRDs based solely on photometric criteria might include non-AGN sources such as Galactic brown dwarfs \citep{Langeroodi_2023}, 
thereby possibly overestimating the AGN abundance. 
\cite{Greene_2024} reported the identification success rate of AGNs among LRDs in the UNCOVER field to be approximately 60\%.
\citet{Kokorev_2024} studied LRDs based on the photometric data from multiple JWST survey areas, using color selection conditions 
provided by \citet{Greene_2024}, which effectively remove contaminants of Galactic brown dwarfs.
To date, other types of low-z interlopers, such as Balmer break galaxies, for LRDs have not been reported via spectroscopic studies.

In contrast, spectroscopic data particularly with measurements of broad H$\alpha$ emission facilitate confirmation of the AGN presence and 
a more accurate determination of $L_{\rm bol}$. 
This method adopts an empirical relationship derived from local AGN observations \citep{Greene_Ho_2005}. 
For LRD sources with detected H$\alpha$ emissions, the continuum fluxes at 5100~{\AA} calculated through the two methods 
yield ratios of $L_{5100, \rm H\alpha}/L_{5100, \rm c} \simeq 1.2 \pm 0.2$ \citep{Greene_2024}.
This result supports the scenario that the continuum emission at 5100~{\AA} is substantially AGN-origin, not from dust-reddened 
stellar continuum of the host galaxy (i.e., $f_L \ll 1$).

One limitation in the work by \cite{Greene_2024} is the use of low-resolution PRISM spectra for analyzing H$\alpha$ emissions, 
which complicates the decomposition of broad and narrow H$\alpha$ emissions in some cases.
To address this challenge, we can use an empirical relationship obtained through higher-resolution spectroscopic observations by \cite{Matthee_2024}. 
This relationship examines the flux ratio between narrow and broad H$\alpha$ emissions in LRDs; namely, 
the ratio $F_{\rm H\alpha,broad}/F_{\rm H\alpha,tot}$ exceeds 0.6 with a positive rest-frame optical spectral index, 
with the ratio approaching unity as the spectral index increases. 
Therefore, the potential systematic errors in estimating broad H$\alpha$ line luminosity due to incomplete spectral line decomposition could 
be alleviated by employing this relationship.

Despite these complexities found in current analyses, the central conclusion of our discussion remains valid on a qualitative level.
Nevertheless, the development of more quantitative arguments will benefit from further observational explorations that 
refine the understanding of the AGN characteristics of LRDs and provide a more precise estimate of their cosmic abundance \citep[e.g.,][]{Li_2024}.

\subsection{Multi-messenger Counterparts}\label{sec:radio}
In this study, we propose a scenario where the dust-rich environments within LRDs lead to the emergence 
of rapidly spinning and overmassive BH.
The high spins of these BHs may yield a strong correlation between the presence of relativistic jets \citep[e.g.,][]{Blandford_Znajek_1977}, 
high-energy emissions and particles \citep[e.g.,][]{Dai_2017,Murase_2020}, and transient bursts such as stellar tidal disruption events 
\citep[e.g.,][]{Inayoshi_2024} with early BHs being overmassive compared to the mass correlation observed in the nearby universe.

Multi-wavelength surveys including radio, optical, and X-ray bands have reported that the radio powers (or radio loudness)
of obscured AGNs initially identified by the VLA/FIRST survey as bright radio sources tend to increase with redshifts at $0.5<z<3.5$, 
with jet powers exceeding $P_{\rm jet}\gtrsim 10^{46}~{\rm erg~s}^{-1}$ \citep{Ichikawa_2021,Ichikawa_2023}.
The inferred jet production efficiency calculated from $\eta_{\rm jet}\sim \epsilon_{\rm rad} P_{\rm jet}/L_{\rm bol}$ approaches unity,
implying rapid spins of these nuclear BHs \citep[e.g.,][]{Tchekhovskoy_2011}.
These types of radio AGNs at intermediate redshifts may offer valuable insights into understanding the characteristics of LRDs at higher redshifts, 
thus further supporting our hypothesis of rapid BH spins.

\subsection{Implications to BH growth mechanisms at $z\gtrsim 5$}

The spin of a massive BH is influenced by a range of physical processes including BH mergers and mass accretion.
The coalescence of two non-spinning BHs results in a remnant with a significant spin, approximately $a_\bullet \simeq 0.69$ for 
an equal-mass merger \citep{Berti_2007}, but the spin diminishes if the merging BHs have non-zero, misaligned spins relative to the orbital 
angular momentum \citep[e.g.,][]{Berti_2008}.
Gas accreting onto BHs through a disk aligned with the angular momentum direction of the BH is expected to enhance 
the spin during mass accumulation \citep{Bardeen_1972,Thorne_1974}.
However, chaotic accretion characterized by short-lived episodes with random orientations tends to dampen the BH spin toward
average values of $a_\bullet \simeq 0.2$ \citep[e.g.,][]{King_2008}.

Chaotic accretion and related feeding mechanisms with low angular momentum gas are considered to
promote the efficiency of BH mass growth in the early universe, due to moderate centrifugal support \citep{Eisenstein_Loeb_1995} 
and a lower radiative efficiency as a result of spin down \citep{King_2008}.
However, we show that the majority of the early BH populations growing in dust-rich
environments tend to exhibit high spins with $a_\bullet \simeq 0.9$, corresponding to $\epsilon_{\rm rad}\gtrsim 0.2$.
This suggests that BH growth is likely dominated by prolonged accretion episodes with coherent angular momentum,
consistent with an idea pioneered by \cite{Volonteri_2005}, or a modest degree of anisotropy \citep{Dotti_2013,Dubois_2014}.
Cosmological simulations focused on galaxy assembly suggest that BHs retain high spins through coherent accretion modes.
This occurs once BHs are settled down to the centers of their host galaxies and then the BH spin direction is well aligned with the angular momentum of the hosts \citep{Peirani_2024}.
As a consequence, a significant fraction of AGNs are expected to launch radio jets and influence the BH-galaxy coevolution \citep[e.g.,][see also Section~\ref{sec:radio}]{Beckmann_2024}.

Our conclusion regarding the rapid spins of BHs can be directly testable through future gravitational-wave observations with 
the space-based detectors such as LISA \citep[e.g.,][]{Amaro-Seoane_2023}.
Moreover, if these BHs frequently merge during galaxy coalescences leading to the LRD phase or similar activities observed in 
ultra-luminous infrared galaxies, such events could significantly contribute to a stochastic gravitational-wave background, 
detectable by pulsar-timing array experiments (\citealt{Inayoshi_2018_GWB}; see also \citealt{PTA_Nanograv_2023}).
Therefore, further exploration of the rapidly spinning, overmassive BHs in LRDs is needed in multiple aspects.

\acknowledgments
We greatly thank Luis C. Ho for emphasizing the significance of the Soltan-Paczy\'nski argument and 
for dedicating time to an inspiring discussion during the spring festival holiday.
We also thank Dale D. Kocevski for sharing their SED fit results for the two LRDs detected in X-rays, which are used in Figure~\ref{fig:ox}.
We also wish to thank Xiaoyang Chen, Anna-Christina Eilers, Joseph F. Hennawi, Haojie Hu, Kazumi Kashiyama, Vasily Kokorev, Masafusa Onoue, Elia Pizzati, and Yasushi Suto for constructive discussions. 
K.~Inayoshi acknowledges support from the National Natural Science Foundation of China (12073003, 12003003, 11721303, 11991052, 11950410493), 
and the China Manned Space Project (CMS-CSST-2021-A04 and CMS-CSST-2021-A06). 
This work is also supported by Japan Society for the Promotion of Science (JSPS) KAKENHI (20H01939; K.~Ichikawa).

\bibliography{ref}{}
\bibliographystyle{aasjournal}

\end{document}